\documentclass[jkps,twocolumn,showpacs,showkeys]{revtex4}
\usepackage[pdftex]{graphicx}
\usepackage{amssymb}
\usepackage{amsmath}
\usepackage{bm}
\usepackage{array}
\usepackage{makecell}
\usepackage{xcolor}
\usepackage{booktabs}
\usepackage{multirow}
\usepackage{ulem}

\newlength\figurewidth
\begin{document}

\title[]{Machine learning for the diagnosis of early stage diabetes using temporal glucose profiles}
\author{Woo Seok \surname{Lee}}
\affiliation{Center for Theoretical Physics of Complex Systems, Institute for Basic Science (IBS), Daejeon, 34126}

\author{Junghyo \surname{Jo}}
\affiliation{Department of Physics Education and Center for Theoretical Physics, Seoul National University, Seoul, 08826}

\author{Taegeun \surname{Song}}
\email{tsong@postech.ac.kr}
\affiliation{Department of Physics, Pohang University of Science and Technology, Pohang, 37673}

\date{\today}

\begin{abstract}
Machine learning shows remarkable success for recognizing patterns in data. 
Here we apply the machine learning (ML) for the diagnosis of early stage diabetes, which is known as a challenging task in medicine.
Blood glucose levels are tightly regulated by two counter-regulatory hormones, insulin and glucagon, and
the failure of the glucose homeostasis leads to the common metabolic disease, diabetes mellitus.
It is a chronic disease that has a long latent period the complicates detection of the disease at an early stage. 
The vast majority of diabetics result from that diminished effectiveness of insulin action.
The insulin resistance must modify the temporal profile of blood glucose.
Thus we propose to use ML to detect the subtle change in the temporal pattern of glucose concentration.
Time series data of blood glucose with sufficient resolution is currently unavailable, so we confirm the proposal using synthetic data of glucose profiles produced by a biophysical model that considers the glucose regulation and hormone action. 
Multi-layered perceptrons, convolutional neural networks, and recurrent neural networks all identified the degree of insulin resistance with high accuracy above $85\%$.
\end{abstract}

\maketitle

\section{INTRODUCTION}
\begin{figure}[h]
\includegraphics[width=\figurewidth]{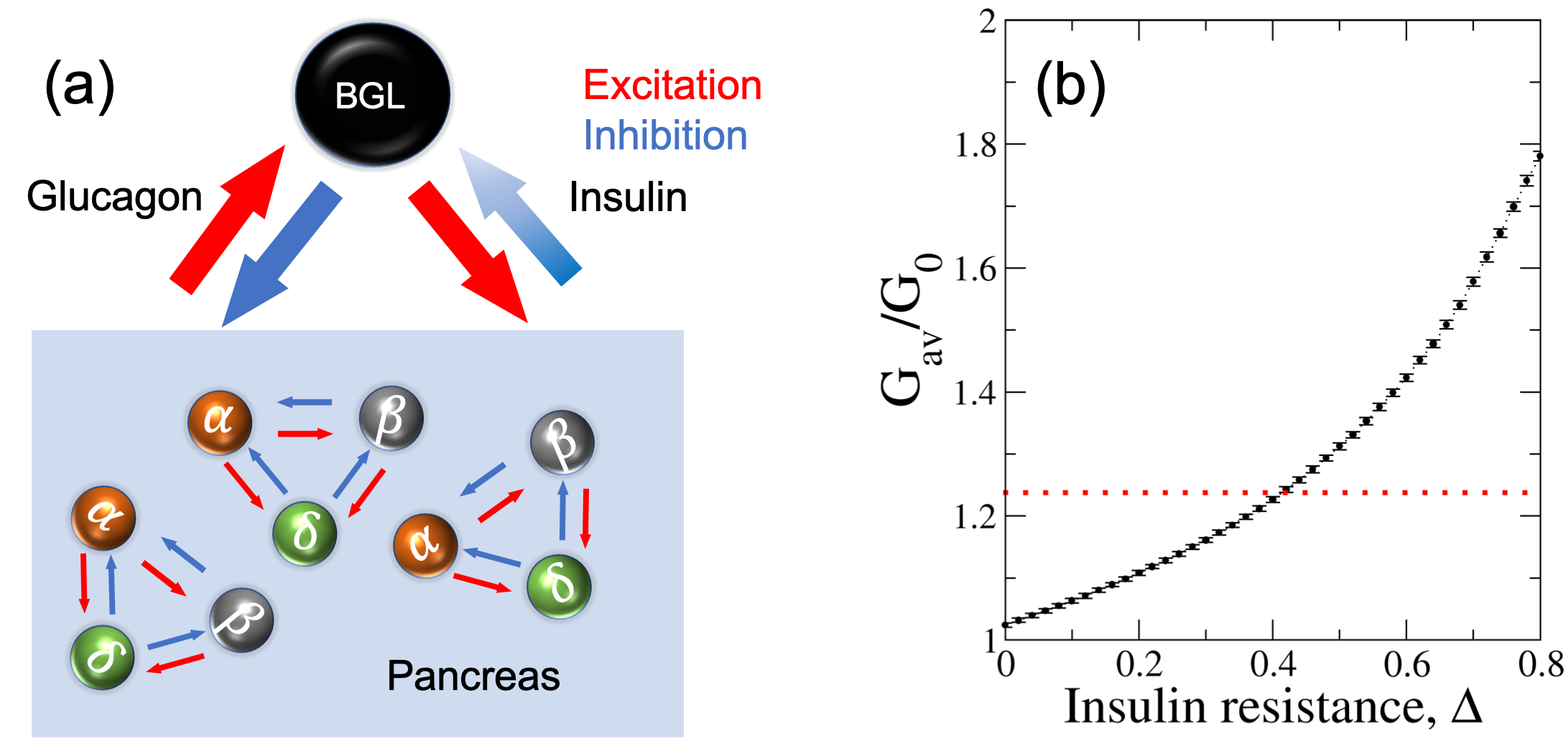}
\caption{(a) Schematic diagram of glucose homeostasis. Blood glucose levels (BGLs) are regulated by insulin and glucagon secreted from endocrine systems. In the pancreas (blue boxed area), each endocrine system consists of three cell types that interact to each other. The blue graded arrow represents the diminished action of insulin, which is insulin resistance. 
(b) Two-hour averaged blood glucose concentration depending on the degrees of insulin resistance. 
The red dotted line represents the standard hyperglycemic threshold of 7.8 mM/$G_0$ $\approx$ 1.24, where $G_0$ = 6.3 mM is a normal glucose concentration. The insulin resistance $\Delta > 0.4$ leads to hyperglycemia.
}
\label{fig1}
\end{figure}

Glucose homeostasis is essential to stably supply fuel to the brain~\cite{GH, GH2}.
Blood glucose levels (BGLs) are tightly regulated by hormones from the endocrine pancreas (Fig.~\ref{fig1} (a)). 
Normal fasting glucose concentration is about 4 mM~\cite{BGL}.
The American Diabetes Association Guideline defines hyperglycemia as $5.6<\text{BGL}<7$ mM.
Severe hyperglycemic ($\text{BGL}> 7.8$ mM average at 2 h fasting) is defined as diabetes mellitus (DM)~\cite{ada_diagnosing}. 
This chronic disease contains long-term damage, dysfunction, and failure of diverse organs resulting in complications. 

DM is grouped into three categories based on the origin of metabolic disorders~\cite{ada_classification}.
Type-1 diabetes mellitus (T1DM) results from insufficient production of insulin due to the destruction of insulin-producing endocrine cells by autoimmunity.
An artificial pancreas can help the patients.
Type-2 diabetes mellitus (T2DM) is a result of diminished effectiveness of insulin action even though it is produced as normal.
T2DM comprises 90\% of the total DM patients. 
Gestational diabetes is a temporary condition in women who develop hyperglycemia during pregnancy. 

Insulin resistance is a key component of health monitoring for a few decades.
The incidence of T2DM is closely related to obesity; in which people, about $60 - 75~\%$ of the population who have body mass index (BMI) $\leq 25~\text{kg}/\text{m}^2$ avoids DM~\cite{seidell2000,seidell1998}.
Insulin resistance is of utmost importance in the pathogenesis of T2DM, hypertension, and coronary heart disease including syndrome X~\cite{reaven1995}.

Here we evaluate machine-learning (ML) as a method to predict development of insulin resistance from the time series of BGLs.
ML has been used to diagnose DM by considering various features of individuals such as age, gender, BMI, waist circumference, smoking, job, hypertension, residential region (rural/ urban), physical activity, and family history of diabetes~\cite{ml2dm1}.
Use of clustering algorithms (linear regression, random forest, k-nearest neighbors, and support vector machine) to evaluate those risk factors can predict whether or not subjects are diabetic.
To date, most ML applications to DM have focused on finding of biomarkers~\cite{ml2dm, ml2dm1}.
In this paper, we provide a novel insight to detect DM development by extracting the increment of insulin resistance, a critical factor of T2DM, from the time trend of BGL.

This idea has not been explored yet, because time series data of BGLs with sufficient temporal resolution are currently not available.
BGLs are regulated by two counter-regulatory hormones, insulin and glucagon, secreted in a pulsatile manner with a $5-10$ min period. 
The signal of fluctuating BGLs can be regarded as an outcome of the balanced response to the insulin and glucagon.
Successful probing of the signal of fluctuating BGLs requires temporal resolution that is fine enough to detect the response of the pulsatile hormones with a shorter time interval than the hormone pulses.
The time resolution of the current state-of-the-art continuous glucose monitoring sensor reaches $5$ min, which is only comparable to the period of hormone pulses.
Therefore, to test our proposal, we use a synthetic data of glucose profiles produced by a biophysical model~\cite{jo2019,jo2017scirep} that considers both glucose regulation and hormone action.

This paper is organized as follows. 
In Section~\ref{sec2}, we briefly introduce the biophysical model that produces time series data of BGLs.
In Section~\ref{sec3}, we explain machine learning methods that we use in this study. 
In Section~\ref{sec4}, we provide results and discussion.

\section{Data preparation of glucose time traces}
\label{sec2}

To produce the data of glucose profiles depending on insulin resistance, we adopt a biophysical model that describes the glucose regulation by endocrine systems~\cite{jo2019,jo2017scirep}. 
Because of the importance of the metabolic disease, diabetes, many biophysical models exist in this field.
Some models describe how glucose stimulates insulin secretion in cellular or organ levels \cite{gsis0, gsis, gsis1}, while other models describe how glucose and insulin regulates each other~\cite{gis2,gis3,gis4,jo2017plosone2}.
Unlike the one-way response model or hormone-level description, the biophysical model we adopt formulates the closed loop between glucose regulation and endocrine systems.

The human pancreas has a few millions of islets, endocrine systems, and each islet consists of $\alpha$, $\beta$, and $\delta$ cells.
Insulin secreted by $\beta$ cells decreases BGLs, whereas glucagon secreted by $\alpha$ cells increases BGLs.
Somatostatin secreted by $\delta$ cells does not directly regulate BGLs, but the three endocrine cell types interact with each other.
The interaction signs between $\alpha$, $\beta$, and $\delta$ cells are very special [Fig. 1(a)].
Depending on the glucose concentration, the endocrine cells show biological rhythms with active/silent phases that lead to corresponding hormone secretion.
The biophysical model describes the rhythmic cellular activities responding to glucose stimuli as phase oscillators modulated by environment~\cite{jo2019,jo2020}.
The model also considers interactions among endocrine cells within islets; these interactions correspond to the coupling in the oscillator model.
The model was used to explain the entrainment of insulin secretion by alternating glucose stimuli in experiment~\cite{jo2017plosone1, jo2017plosone2}.  

In this study, we slightly modified the closed-loop model to consider insulin resistance. 
We use $\sigma \in \{\alpha,\beta,\delta\}$ to represent three types of endocrine cells and $n$ to indicate the islet index.
The activity (or hormone secretion) of $\sigma$ cells in the $n$th islet is denoted by amplitudes $r_{n\sigma}$ and phase $\theta_{n\sigma}$.
The dynamics of the interacting phase oscillators depends on glucose levels $G$:
\begin{eqnarray}
\frac{dr_{n\sigma}}{dt}&=&[f_{\sigma}(G)-r_{n\sigma}^2]r_{n\sigma} \nonumber \\ &&+\sum_{\sigma' \neq \sigma}K_{\sigma\sigma'}r_{n\sigma'}\cos(\theta_{n\sigma'}-\theta_{n\sigma}) \label{amp} \\ 
\frac{d\theta_{n\sigma}}{dt}&=&\omega_{n\sigma}-\mu_{\sigma}(G) \cos(\theta_{n\sigma}) \nonumber\\ &&+\sum_{\sigma'\neq \sigma}K_{\sigma\sigma'}\frac{r_{n\sigma'}}{r_{n\sigma}}\sin(\theta_{n\sigma'}-\theta_{n\sigma})\label{pha}. 
\end{eqnarray}
Here the model describes glucose-dependent amplitude modulations with sigmoidal functions of $f_{\sigma}(G)$ and phase modulations with linear functions of $\mu_{\sigma}(G)$ (\cite{APfunctions} for their specific functional forms).
The spontaneous oscillations of the cellular activities have angular frequencies, $\omega_{n\sigma} \sim \mathcal{N}(\omega_0,0.1)$, which follows a normal distribution with a mean value of $\omega_0=2\pi ~\text{min}^{-1}$ and a standard deviation of 0.1.
The coupling signs between $\alpha$, $\beta$, and $\delta$ cells follow experimental evidence: $K_{\alpha\beta}=K_{\beta\delta}=K_{\alpha\delta}=-1$ and $K_{\beta\alpha}=K_{\delta\alpha}=K_{\delta\beta}=1$. 
Note that islet cells interact only within it; they do not interact with cells located in different islets. 

The total amount of hormone secretion from whole islets is then $\sum_{n} r_{n\sigma}(1+\cos \theta_{n\sigma})$ for glucagon ($\sigma=\alpha$) and insulin ($\sigma=\beta$).
The phase $\theta_{n\sigma}=0$ shows maximal secretion, whereas $\theta_{n\sigma}=\pi$ shows minimal secretion.
Given the fact that insulin decreases the glucose concentration $G$, whereas glucagon increases $G$, so the oscillator model of islets can make a closed loop for glucose regulation:
\begin{eqnarray}
\frac{dG}{dt}&=&\frac{G_0}{N} \sum_{n=1}^{N} r_{n\alpha} (1+\cos  \theta_{n\alpha}) \nonumber \\&&-\frac{G}{N}(1-\Delta)\sum_{n=1}^{N}  r_{n\beta} (1+\cos  \theta_{n\beta}). \label{glu}
\end{eqnarray}
Glucose clearance by insulin is proportional to the present glucose concentration $G$ unlike glucose production by glucagon.
The multiplication of a constant $G_0$ in the glucose production part is included to consider the balance between glucagon and insulin actions at the normal glucose concentration $G_0$.
In this study, we set $G_0=6.3$ mM.
Equations~(\ref{amp})-(\ref{glu}) complete the closed-loop model for glucose regulation in the absence of external glucose stimuli~\cite{jo2019}.
To include the effect of insulin resistance, we introduce an auxiliary parameter $\Delta$ in the glucose clearance part. 
$\Delta$ is a reduction in the effectiveness of insulin action.

As the insulin-resistance parameter $\Delta$ increases, BGLs of $G$ increase [Fig.~\ref{fig1}(b)]. 
In particular, beyond $\Delta=0.4$, the 2-hour averaged BGLs of $G_{\text{av}}$ exceed 7.8 mM, so hyperglycemia is severe.
Therefore, to reproduce early-stage diabetic conditions, we use five groups of $\Delta=\{0, 0.1, 0.2, 0.3, 0.4\}$.
Given $\Delta$, we numerically solved the coupled differential Eqs.~(\ref{amp})-(\ref{glu}) for $N=200$ islets.
Then we took 500 time steps (corresponding to 25 min) with a step size 0.05 min as a sample of glucose time traces.
For each group of $\Delta$, we prepared 2000 samples of the BGL time series for training and 200 samples for testing.
Each sample includes the group label of $\Delta$, which has one-hot encoding (10000, 01000, 00100, 00010, 00001 for $\Delta=0, 0.1, 0.2, 0.3$, and $0.4$, respectively).

\begin{figure}
\includegraphics[width=\figurewidth]{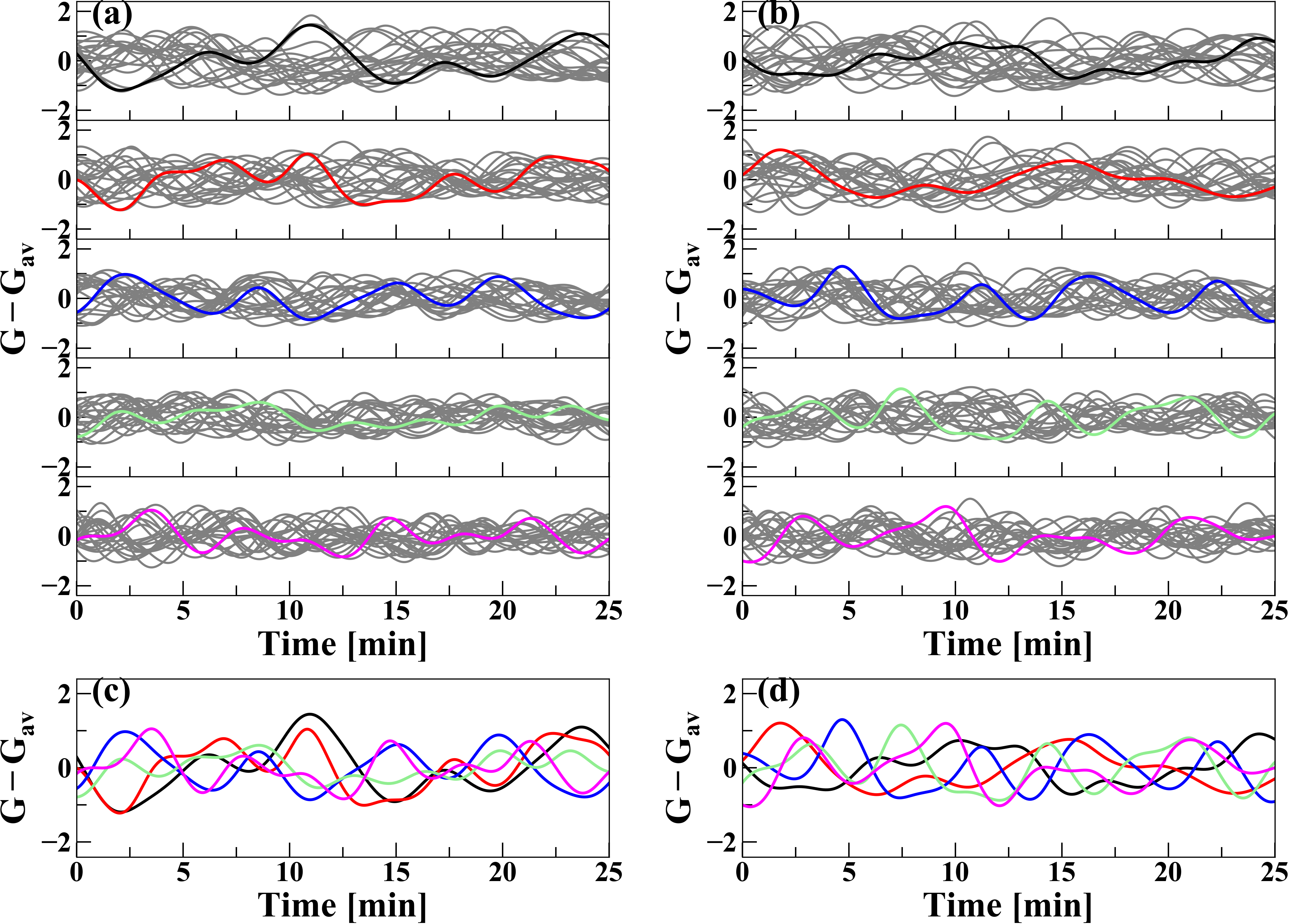}
\caption{Temporal glucose profiles under insulin resistance. Twenty one samples randomly selected (a) from $10000$ training data and (b) from $1000$ test data. Among the gray time series samples, one is highlighted with colors. The five rows have different degrees of insulin resistance $\Delta=\{0, 0.1, 0.2, 0.3, 0.4\}$ from top to bottom. For easy comparison, the colored samples are put together in (c) and (d) for the training and test data, respectively.
}
\label{fig2}
\end{figure}

Different groups of $\Delta$ have a clear feature in the time-averaged value $G_{\text{av}} \equiv 1/\tau \sum_{t=1}^\tau G(t)$ of BGLs for total time step $\tau=500$. However, given real glucose time traces, one cannot judge whether the different $G_{\text{av}}$ results from the different degrees of insulin resistance or simply from individual variations.
Therefore, to avoid this confusion, we focus on the temporal pattern itself rather than the shifted average level by using $G(t) - G_{\text{av}}$.
We produced different samples of $G(t) - G_{\text{av}}$ by solving the model with different initial conditions or by randomly selecting different time windows from full time traces (Fig.~\ref{fig2}).
The temporal patterns for different $\Delta$ are not apparent to the eye.

\section{Pattern recognition of machine learning} \label{sec3}
Temporal pattern classification is one of the most challenging problems in ML~\cite{Esling2012} with a wide range of applications in human activity recognition~\cite{Bevilacqua2018}, electroencephalogram (EEG) classification~\cite{Craik2019}, and speech recognition~\cite{Hinton2012, Mohamed2012}. 
Here, we consider four different neural networks for their ability to classify insulin resistance from the synthesized time traces of BGLs. 
First, we consider a shallow neural network (ShallowNet) as a control for comparison with more sophisticated network models.
Second, we use a fully-connected deep neural network called a multilayer perceptron (MLP); it is the most basic structure for deep learning. 
Third, we use a convolutional neural network (CNN) because it has been very successful in recognition of spatial and temporal patterns~\cite{Krizhevsky2012, Fawaz2019}.
Finally, we also use a recurrent neural network (RNN)~\cite{Sherstinsky2018}; RNNs were originally specialized for temporal data by considering recurrent flows in a network.

Our task is a supervised learning with inputs of temporal glucose traces, $G(t) - G_{\text{av}}$, and outputs of five labels $\Delta$ for insulin resistance.
Thus we assigned $500$ nodes for the input layer, and 5 nodes for the output layer.
We adopt the ReLU (Rectified Linear Unit) as a basic activation function except for the output layer~\cite{LSTM_activation}. 
For the output layer, we use a softmax function to obtain probabilistic predictions as $(p_1,p_2,p_3,p_4,p_5)$.
For example, if $\mathrm{argmax_k}(p_k)=1$, we conclude that the corresponding time trace is the first group, which has $\Delta=0$.
We use the Adaptive Moment Estimation (Adam) algorithm for the optimization of learning~\cite{Kingma2014}.
Now we specify the network structures that we used in this study.

\vspace{0.2cm}
\textbf{ShallowNet}.
The ShallowNet consists of two hidden layers of (1024, 256) nodes for each layer.

\vspace{0.2cm}
\textbf{MLP}. 
The MLP has eight hidden layers of (256, 256, 512, 512, 512, 256, 128, 64) nodes, and every node in a layer is fully connected to every node in their adjacent layers.

\vspace{0.2cm}
\textbf{CNN}.
The CNN for classification is usually composed of two parts. 
The first part is composed of convolution operations that extract features from input data. 
The second part takes the features extracted by the convolution layer and feeds them into the MLP for classification.
The convolution layer consists of a set of trainable filters. 
Each filter convolves across the width and height of input data, and generates convolution outputs. The outputs can be interpreted as a filtered input data.
Therefore, optimizing the filter to extract relevant features from data is a crucial step for CNN.
If one uses many filters, one can extract multiple features.
These convolution processes generate a multi-dimensional feature map, which becomes the input for the MLP that combines all the processed features and finally predicts the classification of input data. 

We prepared two different types of data encoding: 1-dimensional (1D) and 2-dimensional (2D) inputs. 
For 1D input, the feature map is generated by convolution only in the temporal space.
The CNN for 1D input has five convolutional layers, with 100, 100, 200, 200 and 100 filters, respectively, and filter sizes of 10, 10, 10, 3 and 3, respectively. 
A CNN is specialized for 2D image recognition, so we reshaped the 1D temporal data of $(1\times 500)$ into 2D arrays of ``images'' such as $(2\times 250)$, $(5\times 100)$, or $(10\times 50)$. 
Given a sequence $[x_1, x_2, \cdots, x_N]$, the reshaping of ($D\times M$) changes the 1D sequence to $\big[[x_1, x_2, \cdots, x_M]$, $[x_{M+1}, x_{M+2}, \cdots, x_{2M}]$, $\cdots, [x_{(D-1)M+1}, x_{(D-1)M+2}, \cdots, x_{DM}]\big]$.
This 2D reshaping may be able to capture internal structures such as periodicity in the original time traces. For the 2D input, the CNN has also five convolutional layers of 100, 100, 200, 200 and 100 filters respectively, all size (3, 3).
Both 1D and 2D CNNs have the same fully connected MLP parts with four hidden layers with (200, 100, 50, 50) nodes.

\vspace{0.2cm}
\textbf{RNN}.
The RNN is designed to process sequential data. 
Arrays of input data are continuously fed into the RNN, and the activation of each node is transferred to directly connected nodes with recurrent flows.
Thus the RNN can naturally consider the order of time traces. 
Here we use three types of RNNs: (i) vanilla RNN~\cite{Sherstinsky2018}, (ii) long short-term memory (LSTM)~\cite{Sherstinsky2018,Gers1999} and (iii) gated recurrent unit (GRU)~\cite{Cho2014}.
Vanilla RNN is the simplest RNN, and LSTM and GRU consider the memory effect of temporal data. 
Vanilla RNN has three layers with 250 input, 200 hidden, and 50 output nodes for recurrent flows.
For the vanilla RNN, we did not use the 1D input of $(1\times 500)$ because it corresponds to a ShallowNet with a single hidden layer of 200 nodes.
The LSTM and GRU have gated memory cells such as LSTM unit or GRU. 
As for the 2D CNN, we considered various input shapes to test memory effects in the LSTM and GRU networks.

\section{Results and discussions}
\label{sec4}
For the learning, we used the Keras Python using the TensorFlow backend~\cite{keras,tensorflow}.
The learning of this simple task did not take much computation time, usually less than a few tens of minutes.
We examined the diagnosis accuracy (Table ~\ref{table1}), which was measured by the fraction of correct prediction of the degrees of insulin resistance among the tested glucose profiles.
The overall accuracy ranged from 70\% to 90\%; MLP showed the best accuracy.
The accuracy depended on the number of network parameters. 
The numbers of parameters (in millions) were approximately 1.99 (MLP), 0.99 (2D CNN), 0.87 (LSTM), 0.66 (GRU), and 0.26 (vanilla RNN). 

In the analysis of the 2D reshaped data encoding, 2D CNN showed invariant results for different reshaping, whereas GRU and LSTM showed diminished accuracies as the segment length was decreased (Table ~\ref{table2}). This trend can may be a results of the finite filter size of CNN and the memory effect of RNN.
We have also examined a different segmentation rule that changes $[x_1, x_2, \cdots, x_N]$ to $\big[[x_1, x_2, \cdots, x_M]$, $[x_2, x_3, \cdots, x_{M+1}]$, $\cdots, [x_{N-M+1}, x_{N-M+2}, \cdots, x_N]\big]$, but it achieved a negligible increase (1-5\%) in accuracy.

Real glucose profiles include intrinsic and measurement noise.
Thus, to examine the noise effect, we added white noise into our synthesized glucose data, and confirmed that our diagnosis was robust up to about $10~\%$ fluctuation in BGL.

In this study, we checked whether machine learning could detect the patterns of BGL under insulin resistance.
The temporal change of BGL results from the balanced response to the counter-regulatory hormones, insulin and glucagon.
Thus the ineffective action of insulin, called insulin resistance, should affect the BGL profile. 
Therefore, we simulated the glucose profiles under insulin resistance by using a biophysical model for the glucose regulation, and confirmed that the subtle change of glucose profiles under insulin resistance could be recognized by various machine-learning methods.
This demonstrates a great potential of the machine learning approach for the diagnosis of early stage diabetes

A continuous-glucose-monitoring (CGM) system has been widely used for mainly T1DM as a closed-loop artificial pancreas with insulin pumps ~\cite{CGM}. 
The recent development of low-cost CGM has revolutionized CGM usage towards wearable minimally-invasive CGM sensors~\cite{CGM_DM, CGM_AF}.
Such an effort will be in conjunction with our proposal to bring the future direction of diabetes management.
In addition to the difficulty of obtaining high-resolution glucose profiles, the high-accuracy of labels is another prerequisite for successful supervised learning.


\begin{acknowledgments}
This research was supported by the Basic Science Research Program through the National Research Foundation of Korea (NRF), funded by the Ministry of Education, NRF-2019R1F1A1052916 (J.J.), and funded by the Ministry of Science, ICT $\&$ Future Planning through NRF-2017R1D1A1B03034600 (T.S.). 
\end{acknowledgments}



\begin{thebibliography}{}
\bibitem{APfunctions} We set amplitude modulation functions $f_{\alpha}(G)=[1-\tanh((G-G_0)/\delta G)]/2$, $f_{\beta}(G)=[1+\tanh((G-G_0)/\delta G)]/2$,  $f_{\delta}(G)=[1+\tanh((G-G_0+\Delta G)/\delta G)]/4$, and phase modulation functions $\mu_{\alpha}(G)=\mu\cdot (G-G_0)$, $\mu_{\beta}(G)= \mu_{\delta}(G)=-\mu \cdot (G-G_0)$. 
Here, we used the parameters: $G_0=6.3~\text{mM}$, $\Delta G=4.5~\text{mM}$, $\delta G=1.8~\text{mM}$, and $\mu =0.1~\text{mM}^{-1}$.
\bibitem{LSTM_activation} {We used a tangent hyperbolic activation function for the LSTM given input shape $(100\times 5)$, since its long segment dimension causes weight exploding when we used ReLU.}
\end{thebibliography}

%

%


\begin{table*}
\begin{ruledtabular}
\begin{tabular}{@{}cccccccc@{}}
\toprule
\multirow{2}{*}{Accuracy (\%)}                          & \multirow{2}{*}{ShallowNet} & \multirow{2}{*}{MLP} & \multicolumn{2}{c}{CNN} & \multicolumn{3}{c}{RNN} \\
                                           &                             &                      & (1D)         & (2D)         & (vanila)  & (LSTM) & (GRU)   \\ \midrule
\hline
\multicolumn{1}{c|}{$\Delta=0.0$} & 92.5                        & 89.0                 & 79.0   & 84.0       & 88.0    & 82.0  & 85.0  \\
\multicolumn{1}{c|}{$\Delta=0.1$} & 70.5                        & 80.5                 & 63.5   & 78.0       & 70.5    & 81.0  & 80.0  \\
\multicolumn{1}{c|}{$\Delta=0.2$} & 84.5                        & 89.5                 & 65.0   & 89.5       & 76.0    & 83.5  & 81.5  \\
\multicolumn{1}{c|}{$\Delta=0.3$} & 94.0                        & 94.5                 & 73.0   & 89.5       & 80.5    & 84.5  & 89.0  \\
\multicolumn{1}{c|}{$\Delta=0.5$} & 95.0                        & 96.5                 & 84.5   & 91.5       & 92.0    & 99.0  & 97.5  
\\ 
\multicolumn{1}{c|}{Total}   & 87.3                        & 90.0                 & 73.0   & 86.5       & 81.4    & 86.0  & 86.6  \\
\bottomrule
\end{tabular}
\end{ruledtabular}
\caption{Benchmark results of various machine learning methods.
The accuracy for each group $\Delta$ is evaluated with 200 test data corresponding to each group. The total accuracy is evaluated with whole 1000 test data.
}
\label{table1}
\end{table*}

\begin{table*}
\begin{ruledtabular}
\begin{tabular}{@{}cccccccc@{}}
\toprule
\multirow{1}{*}{Input shape}      & \multirow{1}{*}{GRU} & \multirow{1}{*}{LSTM} & \multicolumn{1}{c}{2D CNN} \\
\hline
\multicolumn{1}{c|}{$(2\times 250)$} & 86.6                        & 86.0                 & 85.3    \\
\multicolumn{1}{c|}{$(5\times 100)$} & 84.7                        & 84.5                 & 85.7    \\
\multicolumn{1}{c|}{$(10\times 50)$} & 82.2                        & 84.0                 & 86.5    \\
\multicolumn{1}{c|}{$(25\times 20)$} & 79.5                        & 83.3                 & 84.2    \\
\multicolumn{1}{c|}{$(20\times 25)$} & 80.7                        & 83.9                 & 85.4    \\
\multicolumn{1}{c|}{$(50\times 10)$} & 79.8                        & 82.0                 & 85.5    \\
\multicolumn{1}{c|}{$(100\times 5)$} & 74.6                        & 77.5                 & 82.3   \\ 
\multicolumn{1}{c|}{$(500\times 1)$} & 74.5                        &           73.0     &  -  \\ \bottomrule
\end{tabular}
\end{ruledtabular}
\caption{
Network performance of different 2-dimensional data encoding.
$(D\times  M)$ represents (sement dimension, time steps in each segment).}
\label{table2}
\end{table*}

\end{document}